\documentclass[twocolumn,aps,showpacs]{revtex4-1}
\usepackage{graphicx}
\usepackage{dcolumn}
\usepackage{bm}
\usepackage{mathrsfs}
\usepackage{amsfonts}
\usepackage{amssymb}
\usepackage{amsmath}

\begin{document}

\title{On the Transport Properties of a Quark-Hadron Coulomb Lattice in the
  Cores of Neutron Stars}

\date{15 November, 2012}

\author{Xuesen Na} \email{naxuesen@pku.edu.cn,nxuesen@mail.sdsu.edu}
\affiliation{Department of Astrophysics, School of Physics, Peking
  University,
Beijing 100871, China}

\author{Fridolin Weber} \email{fweber@sciences.sdsu.edu}
\affiliation{Department of Physics, San Diego State University, 5500
  Campanile Drive, San Diego, CA 92182, USA}

\author{Rodrigo Negreiros}
\email{negreiros@fias.uni-frankfurt.de} \affiliation{
Instituto de F\'isica, Universidade Federal Fluminense,
Av. Gal. Milton Tavares de Souza s/n,
Gragoat\'a, Niter\'oi, 24210-346, Brazil}
\affiliation{
  Frankfurt
  Institute for Advanced Studies, Johann Wolfgang Goethe University,
  Ruth-Moufang-Str. 1, 60438 Frankfurt am Main
  Germany}

\author{Renxin Xu} \email{r.x.xu@pku.edu.cn} \affiliation{Department
  of Astrophysics, School of Physics, Peking University,
  Beijing 100871, China}

\begin{abstract}
  Already more that 40 years ago, it has been suggested that because
  of the enormous mass densities in the cores of neutron stars, the
  hadrons in the centers of neutron stars may undergo a phase
  transition to deconfined quark matter. In this picture, neutron
  stars could contain cores made of pure (up, down, strange) quark
  matter which are surrounded by a mixed phase of quarks and
  hadrons. More than that, because of the competition between the
  Coulomb and the surface energies associated with the positively
  charged regions of nuclear matter and negatively charged regions of
  quark matter, the mixed phase may develop geometrical structures,
  similarly to what is expected of the sub-nuclear liquid-gas phase
  transition.  In this paper we restrict ourselves to considering the
  formation of rare phase blobs in the mixed quark-hadron phase. The
  influence of rare phase blobs on the thermal and transport properties of
  neutron star matter is investigated. The total specific heat, $c_V$,
  thermal conductivity, $\kappa$, and electron-blob Bremsstrahlung
  neutrino emissivities, $\epsilon_{\nu,\text{BR}}$, of quark-hybrid
  matter are computed and the results are compared with the associated
  thermal and transport properties of standard neutron star matter. Our
  results show that the contribution of rare phase blobs to the specific
  heat is negligibly small. This is different for the neutrino
  emissivity from electron-blob Bremsstrahlung scattering, which turns
  out to be of the same order of magnitude as the total contributions
  from other Bremsstrahlung processes for temperatures below about
  $10^8$~K.
\end{abstract}


\maketitle

\section{Introduction}\label{sec:intro}

Already many decades ago, it has been suggested that, because of the
extreme densities reached in the cores of neutron stars, neutrons and
protons may transform to quark matter in the cores of such objects
\cite{ivanenko65:a,fritzsch73:a,baym76:a,keister76:a,chap77:a+b,fech78:a}.
Quark matter could thus exist as a permanent component of matter in
the ultra-dense centers of neutron stars (see
\cite{glendenning01:a,glendenning00:book,weber99:book,weber05:a,maruyama07:a}
and references therein). If the dense interior of a neutron star is
indeed converted to quark matter, it must be three-flavor quark matter
since it has lower energy than two-flavor quark matter. And just as
for the hyperon content of neutron stars, strangeness is not conserved
on macroscopic time scales, which allows neutron stars to convert
confined hadronic matter to three-flavor quark matter until
equilibrium brings this process to a halt.

As first realized by Glendenning \cite{glendenning92:a}, the presence
of quark matter enables the hadronic regions of the mixed phase to
arrange to be more isospin symmetric than in the pure phase by
transferring charge to the quark phase in equilibrium with it. The
symmetry energy will be lowered thereby at only a small cost in
rearranging the quark Fermi surfaces. The electrons play only a minor
role when neutrality can be achieved among the baryon-charge carrying
particles. The stellar implication of this charge rearrangement is that
the mixed phase region of the star will have positively charged
regions of nuclear matter and negatively charged regions of quark
matter.

\begin{figure}
  \includegraphics[scale=0.2]{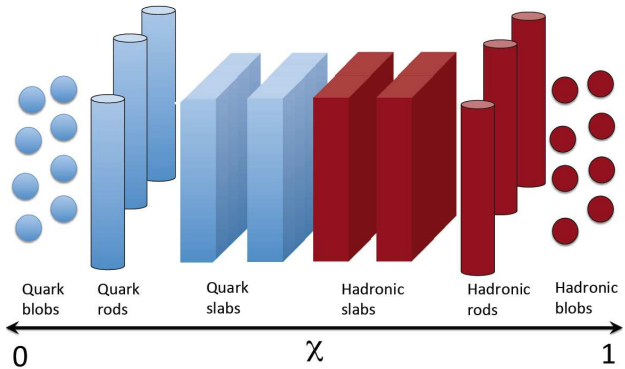}
  \caption{Schematic illustration of possible geometrical structures
    in the quark-hadron mixed phase of neutron stars. The structures
    may form because of the competition between the Coulomb and the
    surface energies associated with the positively charged regions of
    nuclear matter and negatively charged regions of quark matter.}
\label{fig:qh_structures}
\end{figure}

Because of the competition between the Coulomb and the surface
energies associated with the positively charged regions of nuclear
matter and negatively charged regions of quark matter, the mixed phase
may develop geometrical structures (see Fig.\
\ref{fig:qh_structures}), similarly as it is expected of the
subnuclear liquid-gas phase transition
\cite{ravenhall83:a,ravenhall83:b,williams85:a}. This competition
establishes the shapes, sizes, and spacings of the rare phase in the
background of the other in order to minimize the lattice energy
\cite{glendenning92:a,maruyama07:a,glendenning01:a,glen95:a}.

The change in energy accompanied by developing such geometrical
structures is likely to be very small in comparison with the volume
energy
\cite{glendenning92:a,glen01:b,heiselberg92:a,heiselberg95:crete} and,
thus, may not much affect the global properties of a neutron star.
However, the geometrical structure of the mixed phase may be very
important for irregularities (glitches) in the timing structure of
pulsar spin-down as well as for the thermal and transport properties
of neutron stars \cite{glendenning92:a,glen01:b,glendenning00:book}.

To calculate the neutrino-pair bremsstrahlung rates and thermal
properties, we follow the method described in \cite{kaminker99:a} and
\cite{potekhin99:a}, which is commonly used for the calculation of the
neutrino emissivity and thermal conductivity in the crusts of neutron
stars.  These authors considered contributions from electron-phonon
scattering and Bragg diffraction (the static-lattice
contribution). Furthermore, multi-phonon processes and electron band
structure effects are incorporated to obtain more realistic scattering
rates and a better connection between the solid and the liquid gas
phase.  Instead of adopting the analytic fits provided in
\cite{kaminker99:a} and \cite{potekhin99:a}, here we re-calculate the
scattering rates from phonon sums using the method of
\cite{mochkovitch79:a}. There are two main reasons for this. The first
being that, for the crust, the total ion charge is balanced by the
total electron charge. This will be different for the mixed
quark-hadron phase in the core of a neutron star, since electric
charge neutrality is established between the electric charges of the
rare phase, the dominant phase, and the leptons which are present in
both the rare and the dominant phase. The simple relation $n_e=Z n_i$
between electron density and ion density, used to derive the crustal
fit formula in \cite{kaminker99:a,potekhin99:a}, can therefore not be
used to study the quark-hadron Coulomb lattice structure in the core
of a neutron star. The second reason concerns the electric charge
numbers themselves. For mixed phase blobs, they can easily exceed
$Z\sim10^3$, as will be shown in \S~\ref{sec:mixedph}. Charge numbers
that high are obviously not reached in the crustal regimes of neutron
stars \cite{kaminker99:a}, where there is usually no need to consider
atomic nuclei with charges much larger than $Z>56$.

The paper is organized as follows. In Section~\ref{sec2}, we briefly
discuss the modeling of the mixed quark-hadron phase in the cores of
neutron stars and the equations of state of confined hadronic and
quark matter used in this work. In Section~\ref{sec:calc}, we
summarize the formalism for calculating the neutrino-pair
Bremsstrahlung emissivity and the thermal conductivity of rare phase blobs
immersed in hadronic matter. The results are presented in
Section~\ref{sec:result}.

\section{Modeling of the mixed quark-hadron phase in neutron
  stars} \label{sec2}

\subsection{Hadronic matter} \label{sec21}

To compute the particle compositions of the cores of standard neutron
stars, that is, neutron stars without deconfined quark degrees of
freedom, we choose a relativistic lagrangian of the following type
\cite{glendenning00:book,weber99:book},
\begin{eqnarray}
  \mathcal{L} &=& \sum_{B}\bar{\psi}_B\bigl[\gamma_\mu(i\partial^\mu-g_\omega
  \omega^\mu-g_\rho \vec{\tau}\cdot\vec{\rho}_\mu) \nonumber\\
  &-&(m_N-g_\sigma\sigma)\bigr]\psi_B+\frac{1}{2}(\partial_\mu\sigma\partial^\mu
  \sigma-m_\sigma^2\sigma^2) \nonumber\\
  &-&\frac{1}{3}b_\sigma m_N(g_\sigma\sigma)^3-\frac{1}{4}c_\sigma(g_\sigma\sigma)^4-
  \frac{1}{4}\omega_{\mu\nu} \, \omega^{\mu\nu} \nonumber\\
  &+&\frac{1}{2}m_\omega^2\, \omega_\mu\, \omega^\mu+\frac{1}{2} m_\rho^2 \, 
  \vec{\rho}_\mu\cdot
  \vec{\rho\,}^\mu \label{eq:lag}\\
  &-&\frac{1}{4}\vec{\rho\,}_{\mu\nu} \, \vec{\rho\,}^{\mu\nu} + \sum_{\lambda=e^-, \mu^-}
  \bar{\psi}_\lambda
  (i\gamma_\mu\partial^\mu-m_\lambda)\psi_\lambda \, ,
  \nonumber
\end{eqnarray}
where the sum over $B$ sums the baryon species listed in Table
\ref{tab:masses}.  The sum over $\lambda$ accounts for the presence of
relativistic electrons and muons in neutron star matter.  Their masses
are $m_e=0.511$~MeV and $m_\mu=105$~MeV.  The quantities $g_\rho$,
$g_\sigma$, and $g_\omega$ are meson-baryon coupling constants of
$\sigma$, $\omega^\mu$, and
\begin{table}
  \caption{Masses $m_B$, electric charges $Q_B$, spin $J_B$, and third component
    of isospin $I^3_B$ of the  baryons $B$ included in the lagrangian of
    Eq.\ (\ref{eq:lag})
    \cite{glendenning00:book,weber99:book}.}
\label{tab:masses}
\begin{ruledtabular}
\begin{tabular}{cccccc}
$B$  & Symbol & $m_B$ (MeV) &$Q_B$ &$J_B$  &$I^3_B$\\
\hline
$n$ & $m_n$ & 939 &$0$  &$1/2$ &$-1/2$\\
$p$ & $m_p$ &938  &$1$  &$1/2$ &$1/2$\\
$\Lambda$  &$m_\Lambda$ &1115 &$0$  &$0$ &$0$ \\
$\Sigma^+$ & $m_{\Sigma^+}$ &1190 &$1$  &$1$ &$1$ \\
$\Sigma^0$ & $m_{\Sigma^0}$ &1190 &$0$  &$1$ &$0$ \\
$\Sigma^-$ & $m_{\Sigma^-}$ &1190 &$-1$ &$1$ &$-1$ \\
$\Xi^0$ & $m_{\Xi^0}$ &1315 &$0$  &$1/2$ &$0$   \\
$\Xi^-$ & $m_{\Xi^-}$ &1315 &$-1$ &$1/2$ &$-1$ \\
\end{tabular}
\end{ruledtabular}
\end{table}
$\vec{\rho}^\mu$ mesons. Non-linear $\sigma$-meson self-interactions
are taken into account in Eq.\ (\ref{eq:lag}) via the terms
proportional to $b_\sigma$ and $c_\sigma$. The quantities $\vec{\tau}$
and $\gamma^\mu$ denote isospin vectors and Dirac matrices,
respectively, and $\partial^\mu \equiv \partial/\partial x_\mu$
\cite{glendenning01:a,weber99:book}. We have solved the equations of
motion for the baryon and meson fields, which follow from Eq.\
(\ref{eq:lag}), in the framework of the relativistic mean-field
approximation \cite{glendenning01:a,weber99:book}, where the fields
$\sigma$, $\omega$, $\rho$ are approximated by their respective
mean-field expectation values $\bar{\sigma} \equiv
\langle\sigma\rangle$, $\bar\omega \equiv \langle\omega\rangle$, and
$\bar{\rho} \equiv \langle\rho_{03}\rangle$. Two popular
parametrizations labeled G300 and HV have been used
\cite{glendenning85:a,glendenning89:a,glendenning01:a,weber99:book,weber89:a}. Their
parameters are summarized in Table \ref{tab:couplings}.

\begin{figure}
  \includegraphics[scale=0.58]{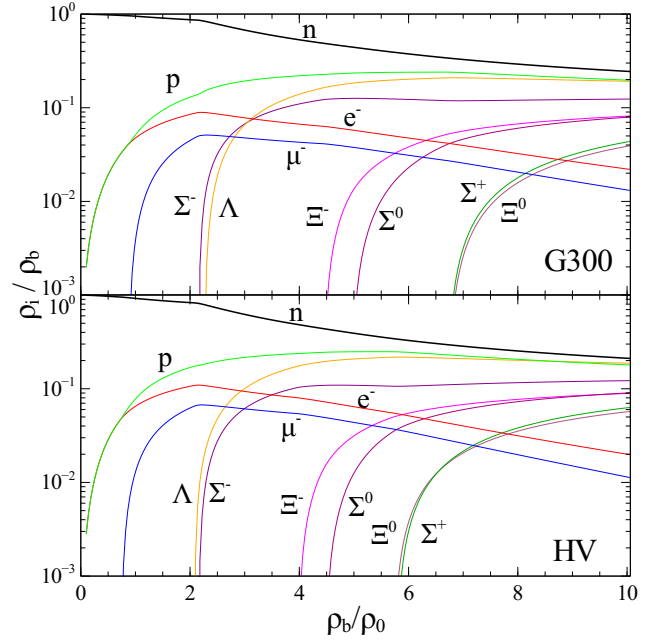}
  \caption{(Color online) Sample baryon-lepton compositions, $\rho_i/\rho_b$, of
    neutron star matter computed for G300 (upper panel) and HV (lower
    panel).}
\label{fig1}
\end{figure}

Neutron star matter is characterized by the conservation of two
charges, electric and baryonic. This feature leads to the chemical
equilibrium condition
\begin{equation}
\mu_i=B_i\mu_n - Q_i\mu_e \, ,
\label{eq:mu_i}
\end{equation}
with $\mu_n$ and $\mu_e$ are the chemical potentials of neutrons and
electrons.  The quantities $B_i$ and $Q_i$ stand for the baryon number
and the electric charge of particles (mesons and baryons) of type $i$.
\begin{table}
  \caption{Relativistic mean-field parametrizations used in this work.}
  \label{tab:couplings}
\begin{ruledtabular}
\begin{tabular}{ccc}
              &\multicolumn{2}{c}{Parametrizations}  \\
\cline{2-3}
Coupling constants   & HV & G300 \\
\hline
$g_\sigma$ & $8.7982$ & $9.1373$ \\
$g_\omega$ & $9.1826$ & $8.6324$ \\
$g_\rho$ & $9.7145$ & $8.3029$ \\
$b_\sigma$ & $0.00414$ & $0.003305$ \\
$c_\sigma$ & $0.00716$ & $0.01529$ \\
\end{tabular}
\end{ruledtabular}
\end{table}
Equation (\ref{eq:mu_i}) greatly simplifies the mathematical analysis,
since only knowledge of two independent chemical potentials, $\mu_n$
and $\mu_e$, is necessary. The latter are given by
\begin{align}
  \mu_B &= g_{\omega} \bar{\omega} + g_{\rho} \bar{\rho_{03}}
  I^3_B + \sqrt{k_B^2+m_B^{*2}} \,,
  \nonumber \\
  \mu_\lambda &= \sqrt{k_\lambda^2+m_\lambda^2} \, ,
\end{align}
where $m_B^*=m_B - g_{\sigma B}\bar{\sigma}$ denote the effective
medium-modified baryon masses, $k_B$ and $k_\lambda$ are the Fermi
momenta of baryons and leptons, respectively, and $I^3_B$ is the third
component of the isospin vector of a baryon of type $B$. Finally,
aside from chemical equilibrium, the condition of electric charge
neutrality is also of critical importance for the composition of
neutron star matter.  It is given by
\begin{eqnarray}
  \sum_B Q_i \, (2J_B+1) \, \frac{k_B^3}{6 \pi^2} -
  \sum_\lambda \frac{k_\lambda^3}{3 \pi^2} = 0 \, .
\label{eq:electric}
\end{eqnarray}
Figure~\ref{fig1} shows the baryon-lepton compositions of neutron star
matter computed from Eq.\ (\ref{eq:lag}) for the relativistic
mean-field approximation.
The quantity $\rho_i$ in Fig.\ \ref{fig1} stands for the individual
number densities of baryons,
\begin{eqnarray}
  \rho_B = (2 J_B + 1) \, k_B^3 / 3 \pi^3 \, ,
\label{eq:rho_B}
\end{eqnarray}
whose total number density is given by
\begin{eqnarray}
\rho_b \equiv \sum_B \rho_B \, .
\label{eq:rho_baryons}
\end{eqnarray}
The individual number densities of electrons and muons ($i=e^-,
\mu^-$) are given by
\begin{eqnarray}
  \rho_i = 2 k_i^3 / 3 \pi^3 \, .
\label{eq:rho_L}
\end{eqnarray}

The total energy density and pressure of the matter, shown in Fig.\
\ref{fig1}, follow from
\begin{align}
  \epsilon_H&=\frac{1}{3}bm_N(g_\sigma\bar{\sigma})^3+\frac{1}{4}c(g_\sigma\bar{\sigma})^4
  +\frac{1}{2}(m_\sigma\bar\sigma)^2 \nonumber \\
  &+\frac{1}{2}(m_\omega\bar{\omega})^2+\frac{1}{2}(m_\rho\bar{\rho})^2\nonumber \\
  &+\sum_B\frac{1}{\pi^2}\int_0^{k_B}k^2dk\sqrt{k^2+m_B^{*2}}\nonumber \\
  &+\sum_\lambda\frac{1}{\pi^2}\int_0^{k_\lambda}k^2dk\sqrt{k^2+m_\lambda^2} \,,
\label{eq:eosH_e}
\end{align}
and
\begin{align}
  p_H&=-\frac{1}{3}bm_N(g_\sigma\bar\sigma)^3-\frac{1}{4}c(g_\sigma\bar\sigma)^4-
  \frac{1}{2}(m_\sigma\bar\sigma)^2 \nonumber \\
  &+\frac{1}{2}(m_\omega\bar{\omega})^2+\frac{1}{2}(m_\rho\bar{\rho})^2\nonumber \\
  &+\sum_B\frac{1}{\pi^2}\int_0^{k_B}k^2dk\sqrt{k^2+m_B^{*2}}\nonumber \\
  &+\sum_\lambda\frac{1}{\pi^2}\int_0^{k_\lambda}k^2dk\sqrt{k^2+m_\lambda^2}
  \, . \label{eq:eosH_p}
\end{align}

\subsection{Quark matter} \label{sec21b}

To model quark matter, we use the MIT bag model. Up ($u$) and down
($d$) quarks are treated as massless particles while the strange quark
($s$) mass is assigned a valued of $m_s=200$~MeV. First-order
perturbative corrections in the strong interaction coupling constant
$\alpha$ are taken into account
\cite{chodos74:a,chodos74:b,farhi84:a,weber05:a}. The Landau
potentials of up and down quarks are then given by
\begin{align}
  \Omega_u&=-\frac{\mu_u^4}{4\pi^2}\left(1-\frac{2\alpha}{\pi}\right) \,, \\
  \Omega_d&=-\frac{\mu_d^4}{4\pi^2}\left(1-\frac{2\alpha}{\pi}\right) \,,
\end{align}
while for strange quarks we have
\begin{align}
  \Omega_s&=-\frac{1}{4\pi^2}\Bigl\{\mu_s\sqrt{\mu_s^2-m_s^2}(\mu_s^2-\frac{5}{2}m_s^2)
  \nonumber\\
  &+\frac{3}{2}m_s^4 f(\mu_s,m_s)\nonumber\\
  &-\frac{2\alpha}{\pi}\Bigl[3\left(\mu_s\sqrt{\mu_s^2-m_s^2}-m_s^2f(\mu_s,m_s)\right)^2
  \\
  &- 2 (\mu_s^2-m_s^2)^2+3m_s^4\ln^2\frac{m_s}{\mu_s} \nonumber\\
  &+6\ln\frac{\sigma}{\mu_s}\left(\mu_s
    m_s^2\sqrt{\mu_s^2-m_s^2}-m_s^4f(\mu_s,m_s)\right)\Bigr]\Bigr\}
  \nonumber \,,
\end{align}
where $f(\mu,m) \equiv \ln((\mu+\sqrt{\mu^2-m^2})/m)$, and $\sigma$ is
a renormalization constant whose value is of the order of the chemical
potentials \cite{farhi84:a}. In this article we take $\sigma=300$~MeV.
The Landau potentials of electrons and muons are given by
\begin{align}
  \Omega_e&=-\frac{\mu_e^4}{12\pi^2} \,,  \\
  \Omega_\mu&=-\frac{1}{4\pi^2}\Biggl(\mu_\mu\sqrt{\mu_\mu^2-m_\mu^2}(\mu_\mu^2-
    \frac{5}{2}m_\mu^2) \nonumber \\
  &+\frac{3}{2}m_\mu^4\ln\Biggl(\frac{\mu_\mu+\sqrt{\mu_\mu^2+m_\mu^2}}{m_\mu}\Biggr)\Biggr)
  \,.
\end{align}

The condition of chemical equilibrium leads to
\begin{equation}
\mu_d=\mu_s=\mu_u+\mu_e=\mu_u+\mu_\mu \,.
\end{equation}
The partial baryon number densities of the particles is obtained from
$(i=u, d, s, e^-, \mu^-)$
\begin{eqnarray}
\rho_i = -\partial \Omega_i/\partial \mu_i \, ,
\end{eqnarray}
and the total energy density and pressure of quark matter follows from
\begin{eqnarray}
\epsilon_Q&=\sum_i(\Omega_i+\mu_i\rho_i)+B \, ,
\label{eq:e_eosQ}
\end{eqnarray}
and
\begin{eqnarray}
  p_Q=-B-\sum_i\Omega_i \, , \label{eq:P_eosQ}
\end{eqnarray}
where $B$ denotes the bag constant. For $\alpha,m_s\to 0$ one recovers
from Eqs.\ (\ref{eq:e_eosQ}) and (\ref{eq:P_eosQ}) the standard
equation of state of a massless relativistic quark gas, $P=(\epsilon-4B)
/ 3$.

\subsection{Geometric structures in the mixed quark-hadron
  phase}\label{sec:mixedph}

To determine the possible geometric structures in the mixed phase of
quarks and hadrons, we use the Gibbs condition
\begin{eqnarray}
  p_H ( \mu_n , \mu_e, \{ \phi \} ) = p_Q (\mu_n , \mu_e)
\label{eq:gibbs}
\end{eqnarray}
for phase equilibrium between hadronic matter and quark matter
\cite{glendenning92:a}. The quantity $\{ \phi \}$ in Eq.\
(\ref{eq:gibbs}) stands collectively for the field variables
($\bar{\sigma}$, $\bar\omega$, $\bar\rho$) and Fermi momenta ($k_B$,
$k_\lambda$) that characterize a solution to the equations of confined
hadronic matter. We use the symbol $\chi \equiv V_Q/V$ to denote the
volume proportion of quark matter, $V_Q$, in the unknown volume
$V$. By definition, $\chi$ then varies between 0 and 1, depending on
how much confined hadronic matter has been converted to quark matter.
Equation (\ref{eq:gibbs}) is to be supplemented with the condition of
baryon charge conservation and electric charge conservation.  The
global conservation of baryon charge is expressed as
\begin{eqnarray}
  \rho_b = \chi \, \rho_Q(\mu_n, \mu_e ) + (1-\chi) \,
  \rho_H (\mu_n, \mu_e,  \{ \phi \}) \, ,
\label{eq:mixed_rho}
\end{eqnarray}
and the global neutrality of electric charge is given by
\begin{align}
  0 = \chi \ q_Q(\mu_n, \mu_e ) + (1-\chi) \ q_H (\mu_n, \mu_e, \{
  \phi \}) \, .
\label{eq:mixed_charge} 
\end{align}

\begin{figure}
  \includegraphics[scale=0.58]{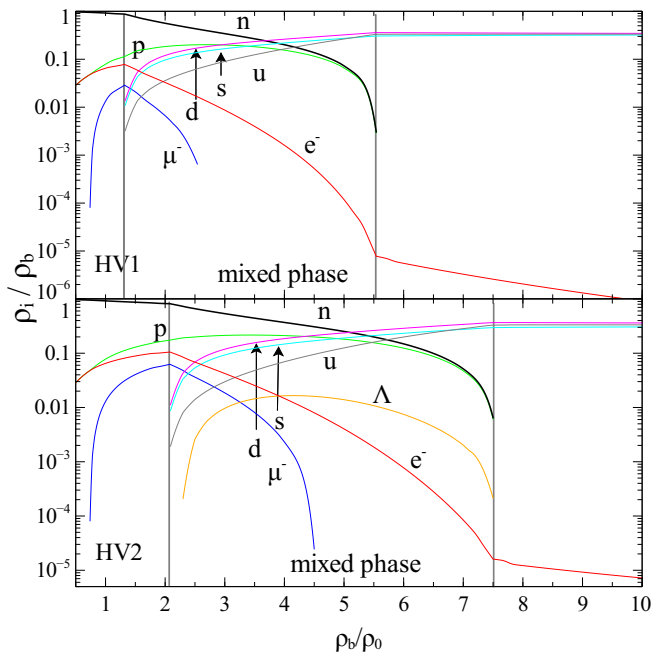}
  \caption{(Color online) Particle composition, $\rho_i/\rho_b$, of
    neutron star matter with mixed quark-hadron phase. The parameter
    sets are HV1 and HV2 (see Table~\ref{tab1}).}
\label{fig21}
\end{figure}

\begin{figure}
  \includegraphics[scale=0.58]{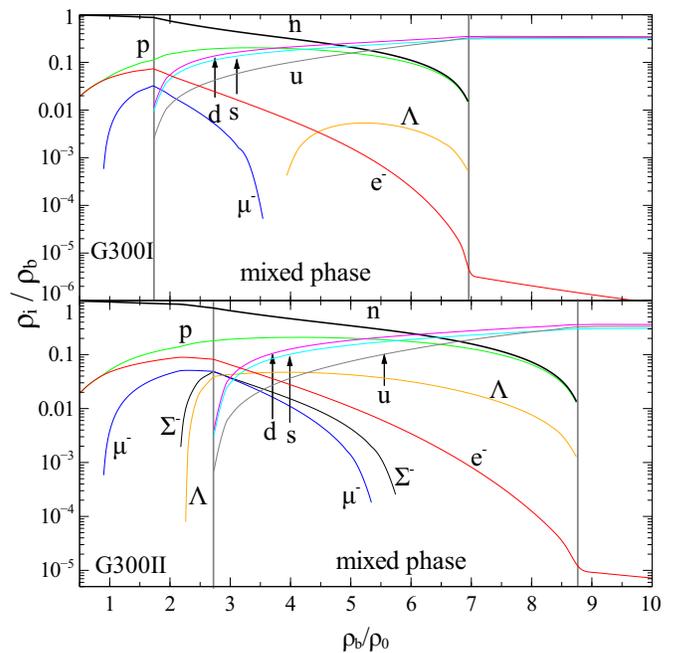}
  \caption{(Color online) Same as Fig.\ \ref{fig21}, but for parameter
    sets G300I and G300II (see Table~\ref{tab1}).}
\label{fig22}
\end{figure}

In Figs.\ \ref{fig21} and \ref{fig22} we show sample compositions of
neutron star matter computed for four different parameter sets (see
Table \ref{tab1}) which allow for the presence of a mixed phase of
quark and hadrons.  One sees that the $\Sigma^-$ population is
strongly suppressed in the mixed quark-hadron phase, since the
hadronic phase carries a net positive charge which disfavors the
presence of the $\Sigma^-$. In contrast to this, the $\Lambda$
particle is electrically neutral so that its presence is not
disfavored by electric charge reasons. As to the population computed
for parametrization G300II, shown in Fig.\ \ref{fig22} (bottom
panel), the threshold of the $\Sigma^-$ is reached before quark
deconfinement
sets in. As soon as quark deconfinement occurs, however, the
$\Sigma^-$ population drops quickly with density while the $\Lambda$
population remains relatively unaffected, enabling hadronic matter to
be positively charged and quark matter to be negatively charged.

As can also be seen from Figs.\ \ref{fig21} and \ref{fig22}, the mixed
quark-hadron phases obtained for the G300I and G300II parametrizations
exists over a broader density range than it it the case for HV1 and
HV2. This feature has it origin in the density dependence of $\mu_n$,
which increases faster with density for the HV1 and HV2 models,
rendering these equations of state stiffer than the G300I and G300II
models for the equation of state.

A larger value for the bag constant, $B$, reduces the pressure of the
quark phase, which too leads to a broader mixed phase region.  It also
follows from Figs.\ \ref{fig21} and \ref{fig22} that the number of
electrons as well as the number of muons drops quickly with density,
since charge neutrality is achieved chiefly among the quarks
themselves.
For values of the strong interaction coupling constant $\alpha \gtrsim
0.25$ electrons disappear from the matter and positrons tend to emerge
in the mixed phase. We therefore consider only $\alpha\le 0.2$.
\begin{table}
  \caption{Parameter sets of the quark-hybrid star models of this work.
    $B$ denotes the bag constant, $\alpha$ is  the
    strong interaction coupling constant.} \label{tab1}
\begin{ruledtabular}
\begin{tabular}{ccc}
Label & Hadronic Phase & Quark Phase \\
\hline
HV1   & HV & $B=110$~MeV~fm${}^{-3}$,~~ $\alpha=0.2$ \\
HV2   & HV & $B=160$~MeV~fm${}^{-3}$,~~ $\alpha=0.1$ \\
G300I & G300 & $B=110$~MeV~fm${}^{-3}$,~~ $\alpha=0.2$ \\
G300II & G300 & $B=160$~MeV~fm${}^{-3}$,~~ $\alpha=0.1$ \\
\end{tabular}
\end{ruledtabular}
\end{table}

\begin{figure}
  \includegraphics[scale=0.58]{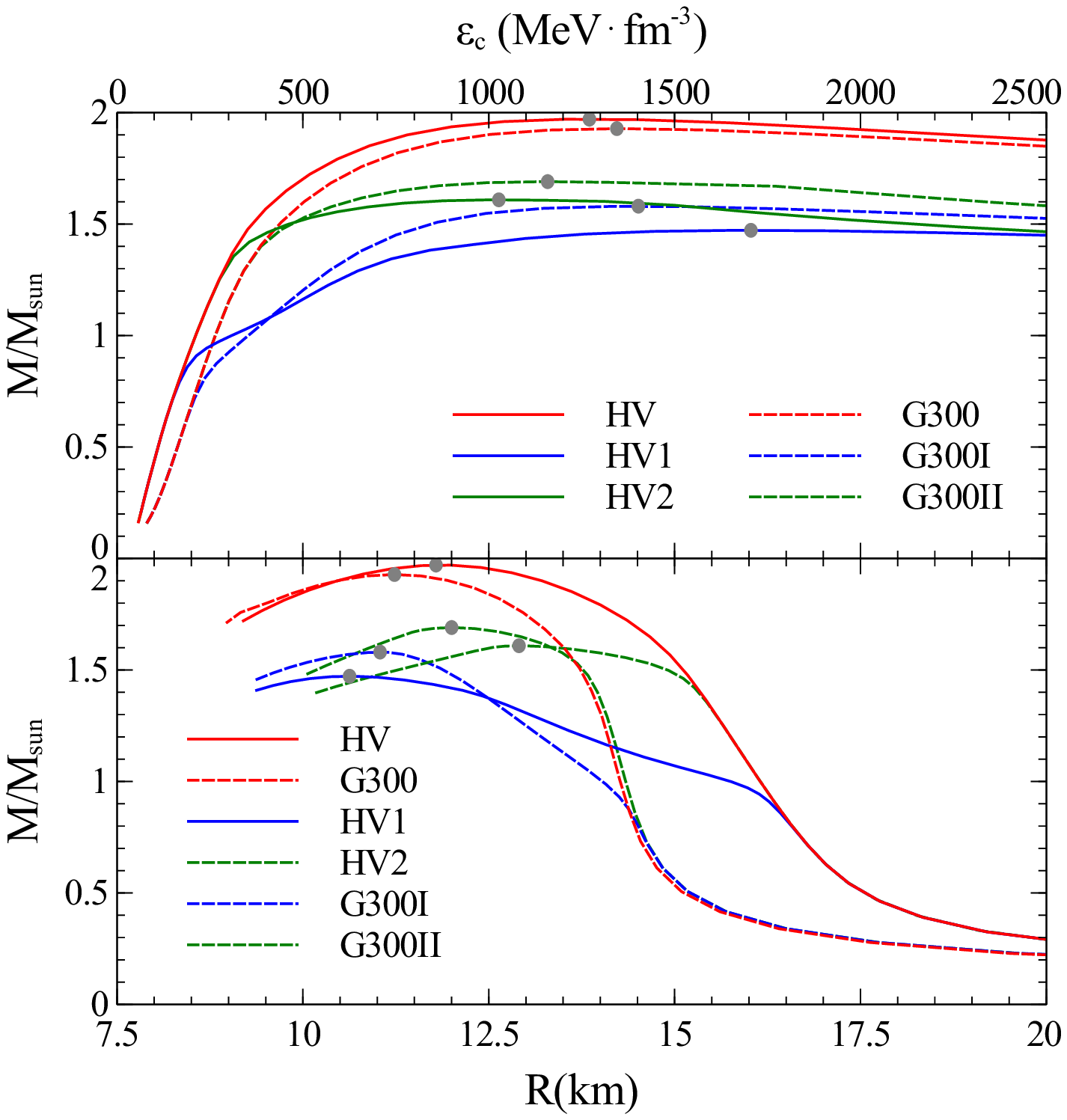}
  \caption{(Color online) Mass-central density (top panel) and
    mass-radius (bottom panel) relationships of neutron stars for the
    parameter sets listed in Table~\ref{tab1}. The solid dots denote
    the most massive neutron star model of each sequence.}
\label{figTOV}
\end{figure}

In Figure \ref{figTOV} we show the masses of neutron stars, whose
compositions are given in Figs.\ \ref{fig21} and \ref{fig22}. The
underlying equations of state are given in Eqs.~(\ref{eq:eosH_e}),
(\ref{eq:eosH_p}), (\ref{eq:e_eosQ}) and (\ref{eq:P_eosQ}), and the
parameter sets are listed in Table \ref{tab1}. The
Baym-Pethick-Sutherland (BPS) model for the equation of state has been
used to model the crusts of these neutron stars \cite{baym71}. The
maximum masses of the quark-hybrid stars computed for HV1 and HV2 are
$1.47 ~ {\rm M}_\odot$ and $1.61~ {\rm M}_\odot$, respectively. For
G300I and G300II we obtain maximum masses of $1.58 ~{\rm M}_\odot$ and
$1.69 ~{\rm M}_\odot$, respectively.  These values are too low to
accommodate the recently discovered heavy pulsar PSR J1614--2230,
whose mass is $M= 1.97\pm 0.04\, {\rm M}_\odot$ \cite{demorest10}. One
possible explanation could be that the high rotation rate of this
neutron star prevents the hadrons in the core of this neutron star
from transforming to quark-hadron matter. This neutron star could thus
be made entirely of confined hadronic matter, whose equation of state
is stiffer than the equation of state of quark-hybrid matter,
supporting high-mass neutron stars. As found in \cite{orsaria12:a},
massive ($\sim 2~{\rm M}_\odot$) non-rotating neutron stars with
extended regions of deconfined quarks and hadrons are comfortably
obtained in the framework of the nonlocal SU(3) Nambu-Jona Lasinio
model. This model is not considered in this paper, however, since our
results are largely independent of the particular microscopic
many-body model chosen to determine the equation of state of bulk
quark-hadron matter.

As pointed out in \cite{glendenning01:a,glendenning92:a} the isospin
restoring force can exploit degrees of freedom made available by
relaxing the requirement of strict local charge neutrality to
neutrality on larger scales and form a positively charged hadronic
matter region with lower isospin asymmetry energy and a negatively
charged quark matter region.  The competition between the Coulomb
interaction and the surface energy will result in a crystalline
lattice of the rare phase in the dominant phase. The size and spacing
of the crystalline lattice is determined by minimizing the total
energy. The situation is similar to the atomic nuclei immersed in a
relativistic electron gas in the crusts of neutron stars.

Depending on density, the embedding of the rare phase in the dominant
phase can lead to different geometric structures, including spherical
blobs, rods and slabs. In what follows, we restrict ourselves to the
discussion of spherical blobs.  For an electrically charge neutral
Wigner-Seitz cell, with spherical blob of radius $r_b$, the Coulomb and
surface energy density can be expressed as
\begin{align}
  \epsilon_C&=2\pi\alpha_e\left[q_H(\chi)-q_Q(\chi)\right]^2 r_b^2 \chi \,
 f_3(x) \,, \\
  \epsilon_S&=3\chi\sigma(\chi)/r_b \,,
\end{align}
where $\alpha_e=1/137$ is the fine structure constant,
$x=\text{min}(\chi,1-\chi)$ is the volume fraction of rare phase,
$f_3(x)=(x-3x^{1/3}+2)/5$ is the function $f_d(x)$ for $d=3$, which
arises from calculating the electrostatic binding energy of the cell
\cite{glendenning01:a}.  Due to theoretical difficulties it is very
hard to estimate the surface tension $\sigma$. Here we follow
\cite{glendenning01:a} and take a gross approximation expression for
the surface tension first proposed by Gibbs \cite{myers85:a}, where
the surface energy is proportional to the difference of the energy
densities of the two phases,
\begin{equation}
  \sigma(\chi)=\eta \, L \, \left[ \epsilon_Q(\chi) -
    \epsilon_H (\chi) \right]
  \label{eq:surfaceten} \,,
\end{equation}
where $\eta$ should be on the order of $\eta\sim\mathcal{O}(1)$, and
we take $L=1$~fm.  Three different values for the constant $\eta$
(i.e., 0.5, 1, and 2) are used in our calculations to investigate the
effects caused by uncertainties in the value of the surface tension.

\begin{figure}
  \includegraphics[scale=0.58]{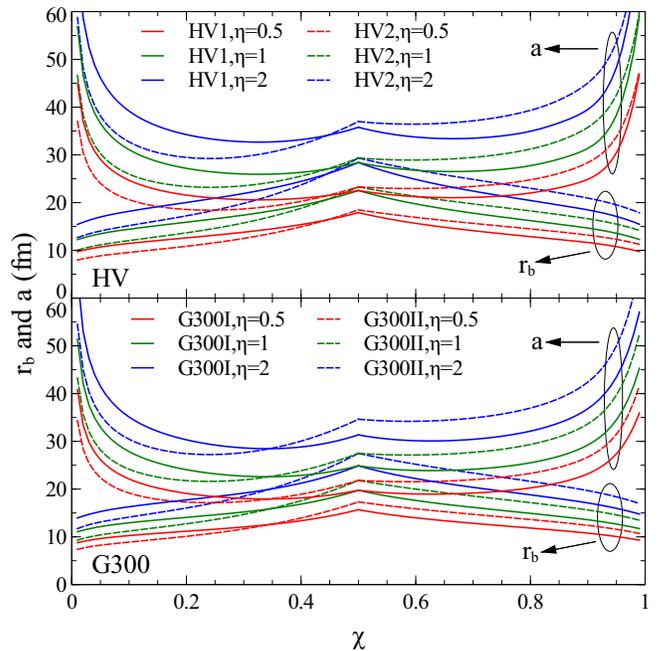}
  \caption{(Color online) Radii of spherical blobs, $r_b$, and
    Wigner-Seitz cells, $a$, as a function of quark volume fraction,
    $\chi$, for different surface tensions, $\eta$, and parameter sets
    (Table \ref{tab1}) of the hadronic lagrangian.}
\label{fig:blobra}
\end{figure}

\begin{figure}
  \includegraphics[scale=0.58]{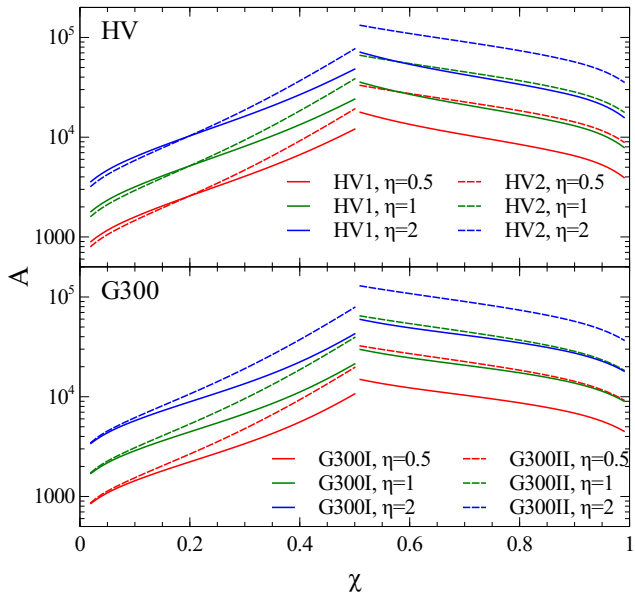}
  \caption{(Color online) Mass number, $A=m_b/m_u$, of spherical blobs
    of rare phase as a function of quark volume fraction, $\chi$.}
\label{fig:blobA}
\end{figure}

\begin{figure}
  \includegraphics[scale=0.58]{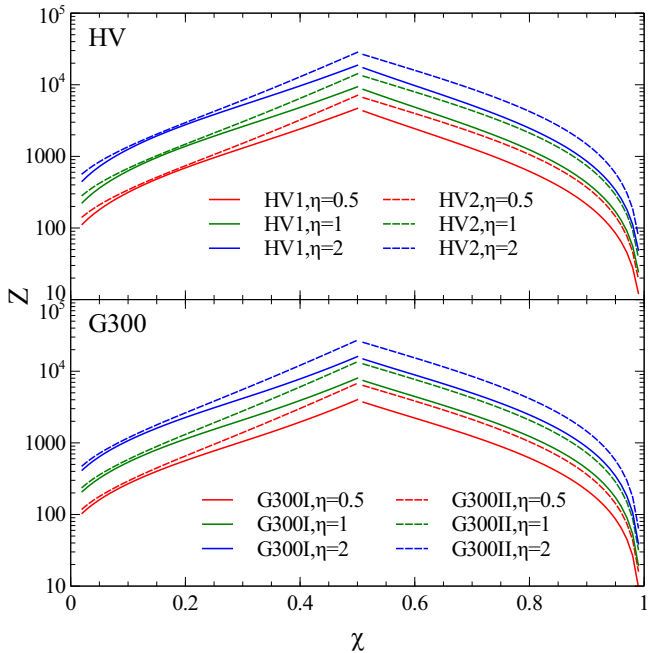}
  \caption{(Color online) Electric charge number, $Z$, of spherical
    blobs of rare phase as a function of quark volume fraction,
    $\chi$.}
\label{fig:blobZ}
\end{figure}

\begin{figure}
  \includegraphics[scale=0.58]{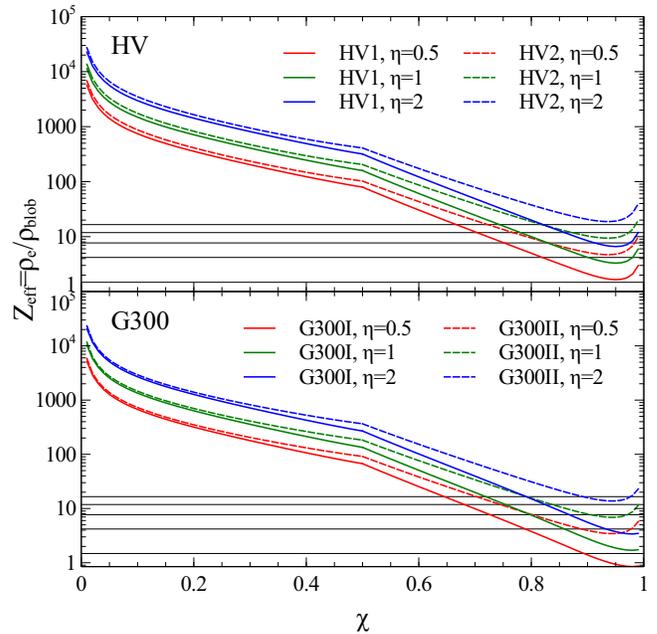}
  \caption{(Color online) Effective charge number, $Z_{\text{eff}}$,
    of spherical blobs of rare phase as a function of quark volume
    fraction. $\chi$. The five horizontal lines at the bottom of each
    panel correspond to $Z_{\text{eff,min}}^{(n)}$ for $n=1,2,3,4,5$
    (see Section~\ref{sec:QBrem}).}
\label{fig:blobZeff}
\end{figure}

Since $\epsilon_C\propto r_b^2$ and $\epsilon_S\propto r_b^{-1}$, it is
possible to minimizing the total energy $\epsilon_C+\epsilon_S$ at
fixed $\chi$ which leads to an equilibrium radius of the rare phase of
blobs inside of Wigner-Seitz cells,
\begin{equation}
  r_b=\left(\frac{3\sigma(\chi)} {4\pi\alpha_e \left[ q_H(\chi) - q_Q(\chi)
      \right]^2 f_3(x)} \right) \,.
\end{equation}
The radii of spherical blobs of the rare phase, $r_b$, and the radii
of Wigner-Seitz cells, $a$, as a function of the quark volume fraction
$\chi$ are shown in Fig.~\ref{fig:blobra}. The radii of spherical
blobs is in the range of 10 to 30~fm. To compare our situation to the
crust we also calculate the charge $Z$ and mass number $A=m_b/m_u$ of
the rare phase blobs, with $m_b$ being the blob mass. In addition we
define an effective electric charge number given by
\begin{equation}
Z_{\text{eff}}=\frac{n_e}{n_b} \,,
\end{equation}
where $n_b$ is the number density of the spherical blobs. The values
of $Z$, $A$ and $Z_{\text{eff}}$ as a function of the quark volume
fraction are shown in Figs.~\ref{fig:blobA}, \ref{fig:blobZ}, and
\ref{fig:blobZeff}.  As can be seen, the charge and mass numbers are
typically one to two orders of magnitude greater than that of the
heaviest stable nuclei, but due to a dramatic drop in electron density
the value of $Z_{\text{eff}}$ can fall near and below
$Z_{\text{eff}}\sim10$ as $\chi\to 1$.  The discontinuities of the
curves in Figure~\ref{fig:blobA} at $\chi=0.5$ are due to the
differences in the hadronic and quark phase densities at $\chi=0.5$.
It is also worth noting that near the edges of the mixed phase region
($\chi\to 0,\, 1$) the volume density of blobs $n_b=3x/(4\pi r_b^3)$
vanishes, but since $f_3(0)=2/5$, the blob radius $r_b$ approaches
a constant $r_b\to r_b(x=0)$. Therefore $Z_{\text{eff}}=n_e/n_b$
diverges at the edges of the mixed phase region.

\section{Thermal and transport properties of quark-hadron phase}
\label{sec:calc}

Next, we turn our interest to the calculation of the thermal and
transport properties of a mixed phase of quarks and hadrons. Knowledge
of these properties is of key importance in order to carry out thermal
evolution simulations of neutron stars with hypothetical quark-hadron
cores and to determine possible astrophysical signatures hinting at the
existence of such matter inside of neutron stars. Our focus here is on
exploring the impact of rare phase blobs on the following properties:
specific heat $c_V$, neutrino emissivity $\epsilon_\nu$, and thermal
conductivity $\kappa$.

From the outset, one might expect that, because of the geometric
structures in the mixed phase, new degrees of freedom are introduced
to the system, which may store additional thermal energy and hence
would increase the specific heat. Due to the scattering of
degenerate electrons and rare phase blobs in the mixed phase, there will
also be an additional contribution to the neutrino Bremsstrahlung and
an additional microscopic entropy production rate which reduces the
total thermal conductivity.

\subsection{Basic physical quantities} \label{sec:basics}

We first introduce some of the basic physical quantities and functions
used in the calculation of the thermal and transport properties of the
quark-hadron phase. The state of the mixed phase with rare phase blobs is
determined by the ion-coupling parameter \cite{kaminker99:a},
\begin{equation}
  \Gamma=\frac{Z^2e^2}{ak_BT} \,,
\end{equation}
where $a=(3/(4\pi n_b))^{1/3}$ is the Wigner-Seitz cell radius, which
is related to the spherical blob radius by $x=(r_b/a)^3$. The quantity
$\Gamma_m=172$ corresponds to the melting point below which a classical
one-component Coulomb crystal ($\Gamma<\Gamma_m$) becomes a Coulomb
liquid ($1<\Gamma<\Gamma_m$) \cite{nagara87:a}.  Therefore the melting
temperature is given by $T_m=Z^2e^2/(ak_B\Gamma_m)$.  Another
important quantity is the plasma temperature,
\begin{equation}
T_p=\frac{\hbar\omega_p}{k_B} \,,
\end{equation}
where $\omega_p=(4\pi Z^2e^2n_b/m_b)^{1/2}$ denotes the plasma
frequency of the spherical blobs, with $m_b$ the mass of the spherical
blobs. For later, it is convenient to introduce the temperature in
units of the plasma temperature, $t_p=T/T_p$.  For temperatures
$t_p\lesssim 1/8$ the vibrations of a Coulomb crystal must be treated
quantum mechanically \cite{kaminker99:a}.

Besides the Coulomb interaction, there are three effects which must be
taken into account to describe the interaction between electrons and
rare phase blobs.  These are the screening of electrons, the shape of the
blob and the effect of thermal vibrations.  The Fourier transform of
the effective electron-blob interaction is given by
\cite{kaminker99:a}
\begin{equation}
  V(q)=\frac{4\pi e\rho_Z F(q)}{q^2\epsilon(q)}e^{-W(q)}  \,,
  \label{eq:Vq}
\end{equation}
where $\rho_Z$ is the blob charge per unit volume $\rho_Z=\chi q_Q$
for $\chi< 1/2$, and $\rho_Z=(1-\chi)q_H$ for $\chi>1/2$, The quantity
$\epsilon(q)$ in Eq.\ (\ref{eq:Vq}) is the static longitudinal
dielectric factor adopted from \cite{jancovici62:a} and $F(q)$ is the
form factor of a blob. For simplicity, we assume a uniform
distribution of the electric charge in the rare phase and use
$F(q)=(3/(qr)^3)[\sin(qr)-qr\cos(qr)]$ \cite{kaminker99:a}.

Thermal vibrations of rare phase blobs are taken into account via the
Debye-Waller factor,
\begin{equation}
  W(q)=\frac{\hbar q^2}{4m_b} \Biggl\langle \frac{\coth
    (\hbar\omega_s/(2k_BT))}
  {\omega_s}\Biggr\rangle_{\text{ph}} \,,
\end{equation}
where $\omega_s$ is the phonon frequency. Here $\langle \ldots
\rangle_{\text{ph}}$ denotes the average over the phonon wave vectors
and polarizations. Throughout this article, we use the method of
\citep{mochkovitch79:a} to compute phonon sums. It is assumed that
there are three polarizations of phonons: two transverse modes with
linear dispersion relations $\omega_k^{(t,i)}=a_i k$ ($i=1,2$), and
one longitudinal mode which is determined through Kohn's sum rule
$\omega_{t,1}^2+\omega_{t,2}^2+\omega_l^2=\omega_p^2$, where
$\omega_p$ denotes the plasma frequency. The two parameters $a_1$ and
$a_2$ are determined by fitting the frequency moments
$u_n\equiv\langle\omega^n\rangle_{\text{ph}}$ of the specified lattice
type.  For a bcc lattice, the frequency moments $u_{-1}=2.7990$ and
$u_{-2}=12.998$ are well known \cite{pollock73:a}, and are used to
obtain $a_1=0.58273$, $a_2=0.32296$. These parameters also produce
first and fourth moments: $\mu_1=0.51106$, $\mu_4=0.201946$ which are
quite close to the exact values of $\mu_1=0.51139$ and $\mu_4=0.203076$
\cite{pollock73:a}.  In the special case of the Debye-Waller factor,
the phonon sum can be fitted very well by the following analytic
formula \cite{baiko95:a},
\begin{equation}
  W(q)=\frac{\alpha_0}{2}\left(\frac{q}{2k_F}\right)^2\left(
    \frac{1}{2}u_{-1}e^{-9.1t_p}+t_pu_{-2}\right) \,,
\end{equation}
where $k_F$ is electron Fermi wave number and $\alpha_0$ is a constant
given in \cite{kaminker99:a}. The latter can be rewritten as
\begin{equation}
  \tilde{\alpha}\equiv\frac{\alpha_0}{2} = \frac{3^{2/3}\pi^{5/6} \hbar^{1/2}}
  {\alpha_e^{1/2}c^{1/2}}
  \frac{n_e^{2/3}}{Zm_b^{1/2}n_b^{1/2}} \, . \label{eq:tildealpha}
\end{equation}
With the help of Eq.\ (\ref{eq:tildealpha}) the Debye-Waller factor
can be written as $W(y^2\tilde{\alpha},t_p)$, where $y=q/(2k_F)$.

\subsection{Neutrino Bremsstrahlung emissivity} \label{sec:QBrem}

Next we turn to the calculation of the scattering of electrons off the
rare phase blobs (that is, electron-blob Bremsstrahlung), which leads to
the generation of neutrino--anti-neutrino pairs according to the
reaction $e^-+(Z,A)\to e^-+(Z,A)+\nu+\bar\nu$. The associated neutrino
Bremsstrahlung emissivity can be written as
\cite{haensel96:a}
\begin{equation}
  \epsilon_\nu=\frac{8\pi G_F^2Z^2e^4C_+^2}{567\hbar^9c^8}(k_BT)^6n_bL \,,
  \label{eq:33}
\end{equation}
where $n_b$ is the number density of the rare phase blobs and
$G_F=1.436\times 10^{-49}$ erg cm${}^3$ is the Fermi weak
coupling constant, $C_+^2\approx 1.675$ \cite{kaminker99:a}, and $L$
is a dimensionless function given by
\begin{equation}
L=L_{\text{ph}}+L_{\text{sl}} \quad \text{or} \quad L=L_{\text{liq}} \,,
\label{eq:34}
\end{equation}
where $L_{\text{ph}}$ accounts for the scattering of electrons off the
phonons of the Coulomb crystal of rare phase blobs, $L_{\text{sl}}$
accounts for Bragg scattering between electrons and the static Coulomb
crystal lattice, and $L_{\text{liq}}$ is for the liquid phase. In the
liquid phase the general expression is obtained through a variational
approach in Born approximation
\cite{festa69:a,haensel96:a,yakovlev96:a},
\begin{equation}
  L_{\rm liq}=\int_0^1dy\frac{S(q)\vert F(q)\vert^2} {y\vert\epsilon(q)\vert^2}
  \left(1+\frac{2y^2}{1-y^2}\ln y\right) \,,
  \label{eq:35}
\end{equation}
where $y=q/(2k_F)$. We follow \cite{kaminker99:a} for the choice of
the ion-ion structure factor $S(q)$, which was fitted in
\cite{itoh83:a,young91:a}.
For a solid phase the phonon contribution is mainly given by umklapp
processes which, in Born approximation, can be written as
\cite{kaminker99:a,yakovlev96:a}
\begin{equation}
  L_{\text{ph}}=\int_{y_0}^1dy\frac{S_{\text{eff}}(y^2\tilde{\alpha},t_p)
    \vert F(y)\vert^2} {y\vert \epsilon(y) \vert^2} \left( 1 +
    \frac{2y^2} {1-y^2} \ln y \right) \,,
    \label{eq:36}
\end{equation}
where the lower integration limit $y_0=(4Z_{\text{eff}})^{-1/3}$
excludes the low-momentum transfers in which the umklapp processes are
forbidden \cite{kaminker99:a}, and the effective static structure
factor $S_{\text{eff}}$ is obtained from the summation of multiphonon
diagrams \cite{baiko98:a,kaminker99:a}. For the parameter sets chosen
in this calculation we always have $Z_{\text{eff}}>1/4$ so that
$y_0<1$. $S_{\text{eff}}$ can be written in terms of a rapidly
decreasing integral \cite{kaminker99:a},
\begin{align}
  S_{\text{eff}}&(y^2\tilde{\alpha},t_p) = 189
  \left(\frac{2}{\pi}\right)^5
  e^{-2W(y^2\tilde{\alpha},t_p)}\nonumber \\
  &\times\int_0^\infty d\xi\frac{1-40\xi^2+80\xi^4}{(1+4\xi^2)^5
    \cosh^2(\pi\xi)}
  \nonumber \\
  &\times\left(e^{\Phi(\xi,y^2\tilde{\alpha},t_p)}-1\right) \,,
\label{eq:S_eff}
\end{align}
where
\begin{equation}
  \Phi(\xi,x,t_p)\equiv x\Biggl\langle\frac{\cos\left(\frac{\xi\omega}
      {t_p}\right)}{\omega\sinh\left(\frac{\omega}{2t_p}\right)}\Biggr
  \rangle_{\text{ph}} \,,
\end{equation}
with $\omega\equiv\omega_s/\omega_p$ denoting the phonon frequency in
units of the plasma frequency.  Similarly to the analysis in
\cite{kaminker99:a}, there exist approximate expressions for
$S_{\text{eff}}$ for the limiting cases where $t_p\ll 1$ and $t_p\gg
1$, which is discussed next. For this purpose, we write the
$x$-independent part of $\Phi(\xi,x,t_p)/t_p$ as
\begin{equation}
  \psi(\xi,t_p)\equiv\frac{1}{t_p}\Biggl \langle \frac{\cos\left(
      \frac{\xi\omega}{t_p}\right)}{\omega\sinh\left(
      \frac{\omega}{2t_p}\right)}\Biggr\rangle_{\text{ph}} \,.
\label{eq:psi_39}
\end{equation}
For $t_p\ll 1$ and $t_p\gg 1$, $\psi(\xi,t_p)$ can be replaced by
\begin{equation}
  \psi(\xi,t_p)\to\begin{cases}
    \psi(0,t_p)\tilde{\psi}(\xi) \, , & t_p\ll 1 \, , \\
    \psi(0,t_p)-\tilde F(t_p)\xi^2 \, , & t_p\gg 1 \, ,
\end{cases}
\end{equation}
where
\begin{equation}
\tilde\psi(\xi)\equiv\lim_{t_p\to 0}\frac{\psi(\xi,t_p)}{\psi(0,t_p)}
\end{equation}
is computed numerically. It is a rapidly decaying
function of $\xi$ and is negligibly small for $\xi\gtrsim 2$.
The function $\tilde F(t_p) \equiv \langle\omega / [2t_p^3\sinh(\omega/2t_p)]
\rangle_{\text{ph}}$ is computed numerically for $1 < t_p < 10^2$.
Asymptotically, $\tilde F(t_p)$ has the form
\begin{equation}
  \tilde F(t_p)=\begin{cases}
    96\left(\frac{1}{a_1^3}+\frac{1}{a_2^3}\right)t_p \, , & t_p\lesssim 1 \, ,\\
    \frac{1}{t_p^2} \, , & t_p\gtrsim 10^2 \,.
\end{cases}
\end{equation}
The function
\begin{equation}
  \psi(0,t_p) = \Biggl\langle \frac{1}{\omega t_p\sinh \left( \frac
      {\omega}{2t_p} \right)} \Biggr\rangle_{\text{ph}} \,,
\end{equation}
can be calculated numerically for a broad range of $t_p$ values, and
can be shown to behave as
\begin{equation}
  \psi(0,t_p)=\begin{cases}
    2u_{-2}-\frac{1}{12t_p^2} \, , & t_p\gtrsim 1 \, , \\
    \pi^2\left(\frac{1}{a_1^3} + \frac{1}{a_2^3}\right)t_p \, ,
    & t_p\lesssim 10^{-3} \,.
\label{eq:psi_43}
\end{cases}
\end{equation}
With the aid of Eqs. (\ref{eq:psi_39}) to (\ref{eq:psi_43}), we can now
derive low and high temperature limits of the effective structure
factor $S_{\rm eff}$ of Eq.\ (\ref{eq:S_eff}). For $t_p\ll 1$ one
obtains
\begin{equation}
  S_{\text{eff}}(x,t_p) =
  189\left(\frac{2}{\pi}\right)^5 e^{-2W(x,t_p)}
  G_{\text{eff}}(x t_p\psi(0,t_p)) \,,
\end{equation}
while for $t_p\gg 1$
\begin{align}
  S_{\text{eff}}(x,t_p) &= 189\left(\frac{2}{\pi}\right)^5
  e^{-2W(x,t_p)+x t_p\psi(x,t_p)} \nonumber \\
  &\times H_{\text{eff}}(xt_pF(t_p))-e^{-2W(x,t_p)} \,.
  \label{eq:H_eff}
\end{align}
Here we have defined
\begin{align}
  G_{\text{eff}}(a) & \equiv\int_0^\infty
  \frac{1-40\xi^2+80\xi^4} {(1+4\xi^2)^5 \cosh^2(\pi\xi)}
  &\times\left(e^{a\tilde{\psi}(\xi)}-1\right) \,,
\end{align}
which obeys
\begin{equation}
  G_{\text{eff}}(a)\simeq\begin{cases}
    e^a \, , & a\gtrsim 10^2 \, ,\\
    \frac{41\pi^5}{181440} a \, , & a\lesssim 0.1 \,.
\end{cases}
\end{equation}
For $0.1\le a\le 10^2$ the value of $G_{\rm eff}$ is obtained
numerically. The quantity $H_{\rm eff}$ in Eq.\ (\ref{eq:H_eff}) is
defined as
\begin{equation}
  H_{\text{eff}}(a)\equiv \int_0^\infty \frac{1-40\xi^2+80\xi^4}
  {(1+4\xi^2)^5\cosh^2(\pi\xi)} e^{-a\xi^2} \, .
\end{equation}
Asymptotically, $H_{\rm eff}$ behaves as
\begin{equation}
  H_{\text{eff}}(a)\simeq\begin{cases}
    \frac{1}{189}(\frac{\pi}{2})^5+0.003690 a  \, , & a\lesssim 0.1 \, ,\\
    (1/2)\pi^{1/2}a^{-1/2} \, , & a\gtrsim 10^3 \,,
\end{cases}
\end{equation}
but its values for $0.1\le a \le 10^3$ need to be computed numerically.

To determined the contributions of the static lattice contribution
(Bragg diffraction) to neutrino Bremsstrahlung, we follow
\cite{kaminker99:a}, who considered band structure effects. We begin
with defining the dimensionless factor $L_{\text{sl}}$,
\begin{align}
  L_{\text{sl}}&=\frac{1}{12Z_{\text{eff}}} \sum_{\mathbf{K}\neq 0}
  \frac{1-y_K^2}{y_K^2}\frac{\vert F(K)\vert^2}{\vert
    \epsilon(K)\vert^2}I(y,t_V)
   e^{-2W(y_K^2\tilde{\alpha},t_p)} \,,
\label{eq:L_SL}
\end{align}
where the sum is over $\mathbf{K}$ values below the electron Fermi
surface, i.e.\ $y_K=K/(2k_F)<1$ and $t_V=\vert V_{\mathbf{K}}\vert
\sqrt{1-y^2}/(k_BT)$ where $V_{\mathbf{K}}$ is given in Eq.\
(\ref{eq:Vq}). The integral $I(y,t_V)$ in Eq.\ (\ref{eq:L_SL}) has
been fitted analytically over a wide range of $t_V$ and $y$ values in
\cite{kaminker99:a}.

In our case, which deals with a solid phase of rare phase blobs immersed in
hadronic matter, the summation of $L_{\text{sl}}$ sometimes consists
only of a few terms.  In the case of a bcc lattice, the condition for
\begin{table}
\caption{Lower bound values $Z_{\text{eff,min}}^{(n)}$.}\label{tab2}
\begin{ruledtabular}
\begin{tabular}{ccccc}
  $Z_{\text{eff,min}}^{(1)}$ & $Z_{\text{eff,min}}^{(2)}$ & $Z_{\text{eff,min}}^{(3)}$
  & $Z_{\text{eff,min}}^{(4)}$ & $Z_{\text{eff,min}}^{(5)}$ \\
  \hline
  $\sqrt{2}\pi/3$ & $4\pi/3$ & $\sqrt{6}\pi$ & $8\sqrt{2}\pi/3$ &
  $5\sqrt{10}\pi/3$ \\
\end{tabular}
\end{ruledtabular}
\end{table}
the summation to have at least $n$ terms (not counting multiplicity)
is $y_{K_n}\le 1$, which translates to a lower limit for
$Z_{\text{eff}}\ge Z_{\text{eff,min}}^{(n)}$. The values of
$Z_{\text{eff,min}}^{(n)}$ for $n=1$ through $n=5$ are shown in
Table~\ref{tab2}.

Therefore, since $Z_{\text{eff}}=Z$ in a lattice of nuclei immersed in
an electron gas, $L_{\text{sl}}$ is guaranteed to have a handful of
terms. However, as we have seen in Section~\ref{sec:mixedph}, the
electron density will drop drastically with increasing baryon number
density in the mixed phase region if $\chi\gtrsim 0.5$ so that
$Z_{\text{eff}}$ drops correspondingly (see Fig.~\ref{fig:blobZeff}
where the first few $Z_{\text{eff,min}}^{(n)}$ are shown).

As will be shown in Section~\ref{sec:result}, for low temperatures of
$T\lesssim 10^8K$ and a quark volume fraction $\chi\gtrsim 0.5$ the
summation consist only of a few terms, and the contribution of the
static lattice oscillates vividly as a function of $\chi$ (see Figs.\
\ref{fQHV} and \ref{fQG300}.)

\subsection{Thermal conductivity} \label{sec:kappa}

To calculate the thermal conductivity of quark-hybrid matter, we
closely follow the formalism outlined in \cite{potekhin99:a}.  In
general, the thermal conductivity of degenerate electrons is expressed
in terms of an effective collision frequency $\nu_\kappa$
\cite{ziman60:book},
\begin{equation}
  \kappa=\frac{\pi k_B^2 T n_e}{3 m_e^*\nu_\kappa} \,, \label{eq:kappa}
\end{equation}
which, in turn, can be expressed in terms of dimensionless Coulomb
logarithms $\Lambda_\kappa$ \cite{yakovlev80:a},
\begin{equation}
  \nu_\kappa^{ei}=\frac{4\pi Z^2 e^4 n_b}{p_F^2 v_F}\Lambda_\kappa  \, ,
  \label{eq:nukappa}
\end{equation}
where $n_b$ is the number density of rare phase blobs in the mixed
phase, and $p_F$ and $v_F$ are the Fermi momentum and Fermi velocity
of relativistic electrons. For a solid phase, the Coulomb logarithms
are calculated variationally in Born approximation
\cite{potekhin99:a,ziman60:book},
\begin{equation}
  \Lambda_{\kappa,{\rm solid}} = \int_{y_0}^1 dy S_\kappa(y)\frac{\vert F(y) \vert^2}
  {y\vert \epsilon(y)\vert^2}\left(1-\frac{x_r^2}{1+x_r^2}y^2\right) \,.
  \label{eq:coullog}
\end{equation}
Here, $x_r\equiv p_F/(m_ec)$ is a relativistic parameter and $y_0$,
$F$, and $\epsilon$ have the same meaning as in
Section~\ref{sec:basics}.  The quantity $S_\kappa$ stands for the
effective static structure factor for thermal conductivity,
\cite{potekhin99:a,baiko98:a}
\begin{equation}
  S_\kappa(y)=S_\sigma(y)+\left(\frac{3}{4y^2} - \frac{1}{2}\right)
  \delta S_\kappa(y) \,,
\end{equation}
where $y=q/(2k_F)$ and $S_\sigma(y)$ are effective structure factors
used to calculate the electric conductivity $\sigma$
\cite{potekhin99:a}.

Instead of using the asymptotic expressions and fitted formulas for
$\Lambda_\kappa$ provided in \cite{potekhin99:a}, which cover $10^{-3}
< t_p <10$ and $0 < \tilde{\alpha} y^2<0.15$ ($\tilde{\alpha}$ is
given by Eq.~(\ref{eq:tildealpha})), here we fully calculate their
values and derive their asymptotic behaviors.  This covers a wider
range of $t_p$ and $\tilde{\alpha} y^2$ values. For this purpose, we
rewrite the relevant integrals given in \cite{potekhin99:a,baiko98:a}
in a form which is similar to $S_{\text{eff}}$ in
Section~\ref{sec:QBrem}, i.e.,
\begin{align}
  S_\sigma(y^2\tilde{\alpha},t_p)&=\int_0^\infty
  d\xi\frac{\pi}{\cosh^2\pi\xi}
  \left(e^{y^2\tilde{\alpha} t_p\psi(\xi,t_p)}-1\right)\nonumber\\
  &\times e^{-2W(y^2\tilde{\alpha},t_p)}\nonumber \,, \\
  \delta S_\kappa(y^2\tilde{\alpha},t_p)&=\int_0^\infty
  d\xi \frac{2\pi(1-2\sinh^2\pi\xi)}{\cosh^4\pi\xi}\nonumber\\
  &\times
  e^{-y^2\tilde{\alpha}t_p\psi(\xi,t_p)-2W(y^2\tilde{\alpha},t_p)} \,,
\end{align}
where $\psi(\xi,t_p)$ denotes the phonon sum function already defined
in Eq.\ (\ref{eq:psi_39}).  Since Bragg diffraction does not
contribute to the thermal conductivity, there is no counterpart to
$L_{\text{sl}}$.

Similar to Section~\ref{sec:QBrem}, asymptotic expressions for
$\psi(\xi,t_p)$ may be used to derive the high and low temperature
limits for $S_\sigma$ and $\delta S_\kappa$. One then obtains
\begin{align}
  S_\sigma&(x,t_p)\xrightarrow{t_p\ll 1} G_\sigma(xt_p\psi(0,t_p))
  e^{-2W(x,t_p)} \nonumber\\
  &\xrightarrow{t_p\gg 1}
  \left(e^{xt_p\psi(0,t_p)}H_\sigma(xt_p\tilde F(t_p))-1\right)
  e^{-2W(x,t_p)} \,, \nonumber \\
  \delta S_\kappa&(x,t_p)\xrightarrow{t_p\ll 1} G_{\delta\kappa}
  (xt_p\psi(0,t_p))e^{-2W(x,t_p)} \nonumber \\
  &\xrightarrow{t_p\gg 1}
  H_{\delta\kappa}(xt_p\tilde F(t_p))e^{xt_p\psi(0,t_p)-2W(x,t_p)} \,,
\end{align}
where $\tilde F(t_p)$ is given in Section~\ref{sec:QBrem} and the functions
$G_{\sigma,\delta\kappa}$ and $H_{\sigma,\delta\kappa}$ are defined as
\begin{align}
  G_\sigma(a)&\equiv \int_0^\infty d\xi\frac{\pi}{\cosh^2\pi\xi}
  \left(e^{a\tilde{\psi}(\xi)}-1\right) \,,
\label{eq:GG} \\
  G_{\delta\kappa}(a)&\equiv \int_0^\infty d\xi
  \frac{2\pi(1-2\sinh^2\pi\xi)}
  {\cosh^4\pi\xi} \, e^{a\tilde{\psi}(\xi)} \,, \\
  H_\sigma(a)&\equiv \int_0^\infty d\xi\frac{\pi}{\cosh^2\pi\xi} \,
  e^{-a\xi^2} \,, \\
  H_{\delta\kappa}(a)&\equiv \int_0^\infty d\xi
  \frac{2\pi(1-2\sinh^2\pi\xi)}{\cosh^4\pi\xi} \, e^{-a\xi^2}\,.
  \label{eq:HH}
\end{align}
The quantity $\tilde{\psi}(t_p)$ is defined in
Section~\ref{sec:QBrem}. The asymptotic limits of Eqs.\ (\ref{eq:GG})
through (\ref{eq:HH}) are given by
\begin{align}
G_{\sigma}(a)&\simeq\begin{cases}
e^a \, , & a\gtrsim 10^2 \, , \\
\frac{2}{3}a \, , & a\lesssim 0.1 \, ,
\end{cases}  \nonumber \\
G_{\delta\kappa}(a)&\simeq \begin{cases}
e^a \, , & a\gtrsim 10^3 \, , \\
\frac{8}{15}a \, , & a\lesssim 10^{-2} \, ,
\end{cases}  \nonumber \\
H_{\sigma}(a)&\simeq \begin{cases}
  \frac{1}{2}\pi^{3/2}a^{-1/2} \, , & a\gtrsim 10^2 \, ,\\
  1-\frac{a}{12} \, , & a\lesssim 0.1 \, ,
\end{cases} \nonumber \\
H_{\delta\kappa}(a)&\simeq \begin{cases}
\pi^{3/2}a^{-1/2} \, , & a\gtrsim 10^3 \, , \\
\frac{2a}{\pi^2} \, , & a\lesssim 10^{-2} \, ,
\end{cases}
\end{align}
For $a$ values outside the ranges listed above the values of $H$ and
$G$ were calculated numerically. Asymptotic approximations for
$S_\sigma$ were used for $t_p\lesssim 10^{-2}$ and $t_p\gtrsim 1$. For
$t_p$ values outside of these intervals the expression for $S_\sigma$
was computed numerically. The asymptotic approximations for $\delta
S_\kappa$ are valid, and have been used, for $t_p$ values in the
intervals $t_p\lesssim 10^{-2}$ and $t_p\gtrsim 10$.

\subsection{Specific heat}\label{sec:cv}

The calculation of the specific heat is much simpler than the
calculation of the neutrino emissivities and of the thermal
conductivity, since the specific heat does not involve scattering
processes. In terms of the phonon sum used in Sections~\ref{sec:QBrem}
and \ref{sec:kappa}, the specific heat density can be written as
\begin{align}
  c_{V,\text{solid}}&=\hbar\omega n_b \frac{\partial}{\partial T}
  \Biggl\langle\frac{3\omega}{\exp\left(\frac{\omega}{2t_p}\right)-1}
  \Biggr\rangle\nonumber\\
  &=\frac{3k_Bn_b}{4t_p^2}\Biggl\langle\frac{\omega^2}{\sinh^2
    \left(\frac{\omega}{2t_p}\right)}\Biggr\rangle \,.
  \label{eq:cv1}
\end{align}
The asymptotic low and high temperature expressions of Eq.\
(\ref{eq:cv1}) are given by
\begin{equation}
  c_{V,\text{solid}}=\begin{cases}
    3k_Bn_b\left(1-\frac{1}{36t_p^2}\right) \, , & t_p\gtrsim 10^2 \, , \\
    72k_Bn_b\left(\frac{1}{a_1^3}+\frac{1}{a_2^3}\right) t_p^3 \, ,
    & t_p\lesssim 10^{-2} \, .
\end{cases}
\end{equation}
For temperatures greater than the melting temperature we adopt
Eq.~(24) of \cite{vanriper91:a} for the specific heat density,
\begin{equation}
  c_{V,\text{liq,gas}}=\begin{cases}
    \frac{3}{2}k_B n_b \, , & \Gamma\le 1\\
    \frac{3}{2}k_B \left( 1+\frac{\log\Gamma}{\log\Gamma_m}\right) \, ,
    & 1<\Gamma\le \Gamma_m \, ,
\end{cases}
\end{equation}
where we use $\Gamma_m=172$ (as in Sections~\ref{sec:QBrem} and
\ref{sec:kappa}) instead of the outdated value $\Gamma_m=150$
\cite{vanriper91:a}.

\section{Results and discussion} \label{sec:result}

\subsection{Contribution to the specific heat} \label{sec:rescv}

In this paper, we have calculated the specific heat stemming from
rare phase blobs immersed in hadronic matter for four different parameter
\begin{figure}[htb]
  \includegraphics[scale=0.58]{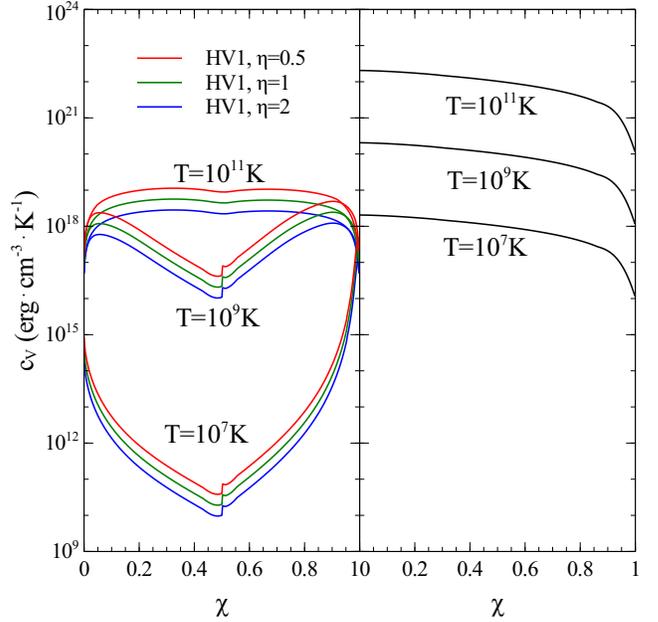}
  \caption{(Color online) Specific heat, $c_V$, of mixed quark-hadron
    phase as a function of quark volume fraction, $\chi$, computed for
    different surface tensions, $\eta$, and temperatures, $T$.  The
    underlying equation of state is HV1. The panel on the left hand
    side shows the contributions of the rare-phase blobs to the
    specific heat.  The curves in the panel on the right hand side
    show the specific heat computed for a standard (no geometrical
    structures) quark-hadron gas.}
\label{fcv1}
\end{figure}
sets (see Table~\ref{tab1}) and three different values ($\eta=0.5$, 1,
2) for the surface tension of rare phase blobs. It is intriguing to compare
these results with the heat capacities of the hadronic and
quark matter phases, weighted by their volume fractions. For this
purpose we compute the specific heat of a Fermi gas of leptons and
baryons from
\begin{align}
  c_V^l&=\frac{k_B^3}{3\hbar^3}\, T\, \sqrt{m_l^2+k_{F,l}^2}\ k_{F,l}
  \,,
  \\
  c_V^b&=\frac{k_B^3}{3\hbar^3}\, T \, \sum_B\, m_B^* \, k_{F,B} \,,
\end{align}
where $m_B^*$ is effective in-medium mass of baryons (see
Section~\ref{sec2}). The specific heat of a quark gas is given by
\cite{iwamoto82:a}
\begin{equation}
  c_V^q=0.6\times 10^{20}\left(\frac{n_e}{\rho_0}\right)^{2/3}
  T_9 ~~ \text{ergs} ~ \text{cm}^{-3}~ \text{K}^{-1} \,.
\end{equation}
The total specific heat of a mixture of quarks and hadrons follows
from $c_V=(1-\chi)c_V^{H}+\chi c_V^{Q}$ where $c_V^H=c_V^b+c_V^l$ and
$c_V^Q=c_V^q+c_V^l$.  Figure~\ref{fcv1} compares the different
contributions to the specific heat with one another.  As can be seen,
the contribution of the rare-phase blobs to the specific heat
(colored lines in Fig.\ \ref{fcv1}) is typically several orders of
magnitude smaller than the specific heat of a standard (no geometrical
structures) quark-hadron gas (solid black lines in Fig.\ \ref{fcv1}).
The small jumps at $\chi=0.5$ in Figure\ \ref{fcv1} are due to the
discontinuity of the rare phase blob mass, $m_b$ at $\chi=0.5$ (see
Fig.\ \ref{fig:blobA}). The parameter sets HV2, G300I and G300II lead
to results very similar to those shown in Fig.\ \ref{fcv1} and are
therefore not shown separately.

\subsection{Neutrino Bremsstrahlung emissivity} \label{sec:resQ}

\begin{figure}
  \includegraphics[scale=0.58]{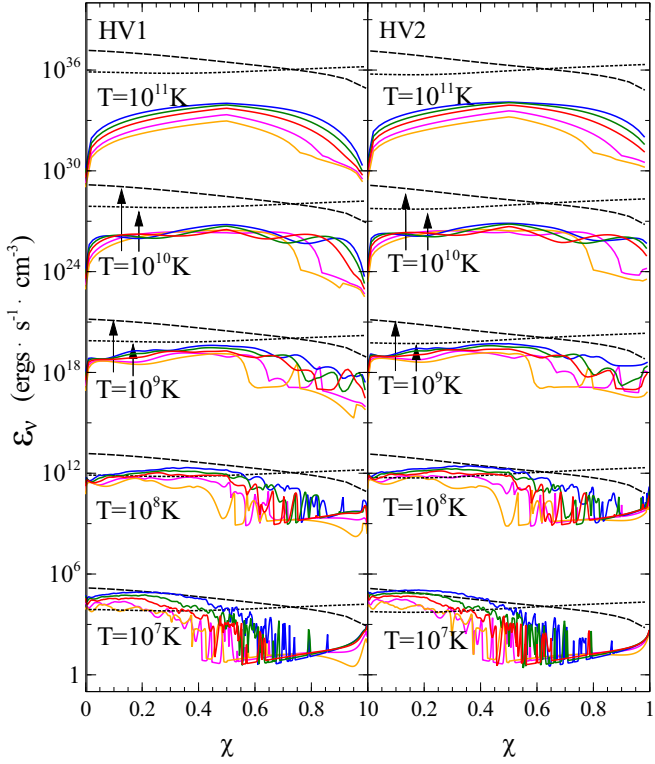}
  \caption{(Color online) Neutrino emissivities, $\epsilon_\nu$, as a
    function of quark volume fraction, $\chi$, at different
    temperatures, computed for parameter sets HV1 and HV2.  The
    colored lines show the total contributions from electron-blob
    Bremsstrahlung for different surface tensions: $\eta=0.1$ (orange),
    $\eta=0.2$ (pink), $\eta=0.5$ (red), $\eta=1$ (green), $\eta=2$
    (blue). The black lines show the total contributions from modified
    nucleon and quark Urca processes (dashed line) and nucleon-nucleon
    and quark-quark Bremsstrahlung processes (dotted line).}
\label{fQHV}
\end{figure}

\begin{figure}
  \includegraphics[scale=0.58]{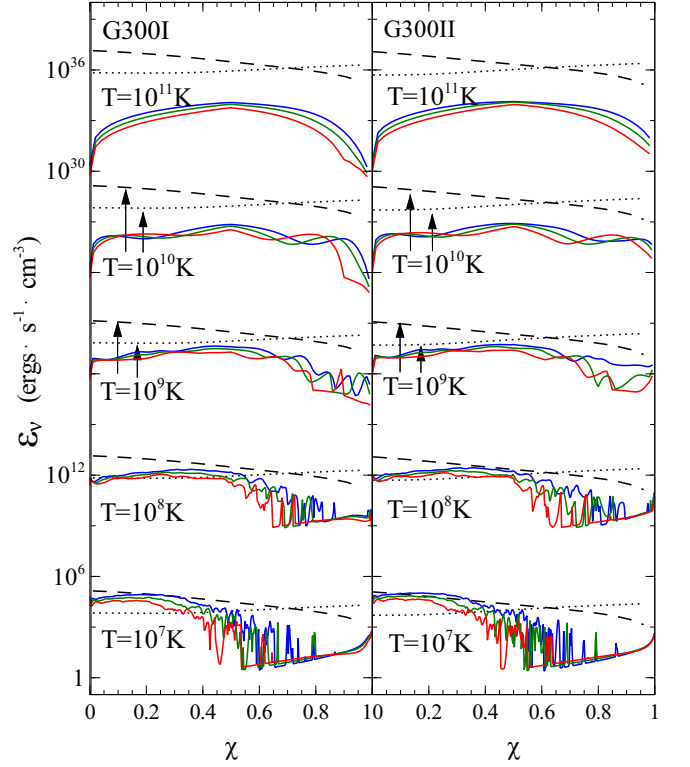}
  \caption{(Color online) Same as Fig.\ \ref{fQHV}, but for parameter
    sets G300I and G300II.  The colored lines show the total
    contributions from electron-blob Bremsstrahlung for different
    surface tensions, $\eta=0.5$ (red), $\eta=1$ (green), $\eta=2$
    (blue). (For the sake of clarity, the curves for $\eta=0.1$ and
    $\eta=0.2$ are not shown.) The black lines shown the total
    contributions from modified nucleon and quark Urca processes
    (dashed line) and nucleon-nucleon and quark-quark Bremsstrahlung
    processes (dotted line).}
\label{fQG300}
\end{figure}

The neutrino Bremsstrahlung emissivities emerging from electron-phonon
scattering and electron-lattice (Bragg diffraction) scattering have
been computed from Eqs.\ (\ref{eq:33}) for the four parameter sets of
Table~\ref{tab1}). The surface tension of rare phase blobs in
Eq.~(\ref{eq:surfaceten}) has been varied again from $\eta=0.5$, 1, to
2.  Figures~\ref{fQHV} and \ref{fQG300} show the contributions of the
rare phase blobs to the neutrino Bremsstrahlung emissivity as
a function of
the quark volume fraction. A range of representative temperatures,
from $T=10^7$~K to $10^{11}$~K has been chosen.  For comparison, we
show the contributions to the neutrino emissivity which comes from the
modified Urca process in hadronic matter,
$\epsilon_{\nu,\text{H,MU}}$, and in quark matter,
$\epsilon_{\nu,\text{Q,MU}}$ \cite{iwamoto82:a}.
Finally, we also show in these figures the emissivities which
correspond to nucleon Bremsstrahlung, $\epsilon_{\nu,\text{H,BR}}$ and
quark Bremsstrahlung, $\epsilon_{\nu,\text{Q,BR}}$
\cite{iwamoto82:a,weber99:book}.  Their total contribution in
quark-hybrid star matter is obtained, for a given quark volume
fraction $\chi$, from $\epsilon_{\nu} = \chi\epsilon_{\text{Q}} +
(1-\chi)\epsilon_{\text{H}}$.  As can be seen from Figs.~\ref{fQHV}
and \ref{fQG300}, the neutrino emissivity from electron-blob
Bremsstrahlung becomes comparable to the emissivities of the modified
Urca process (and other Bremsstrahlung processes) for temperatures $T
\lesssim 10^8$K.

The Bremsstrahlung emissivities oscillate rapidly with $\chi$ for
$T\lesssim 10^9$ and $\chi\gtrsim 0.5$. This is due to Bragg
diffraction, given by the sum in the expression for $L_{\text{sl}}$
(see Eq.\ (\ref{eq:L_SL})).  As mentioned in Section~\ref{sec:QBrem},
the sum of $L_{\text{sl}}$ consists only of a few terms if
$\chi\gtrsim 0.5$. The oscillations are essentially due to the
oscillating values of the individual terms in Eq.\
(\ref{eq:L_SL}). The oscillations are smoothed out when the number of
terms is large.

It also follows from Figs.~\ref{fQHV} and \ref{fQG300} that the
neutrino emissivity of rare phase blob Bremsstrahlung, at low temperatures
but greater quark volume fractions ($\chi\gtrsim 0.5$), becomes
sensitive to the choice of $\eta$.  This feature, however, does not
\begin{figure}[!h]
  \includegraphics[scale=0.58]{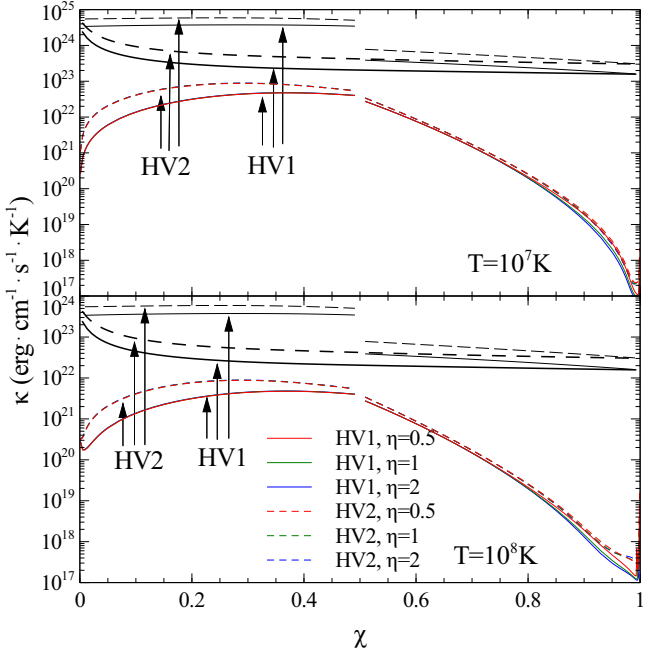}
  \caption{(Color online) Thermal conductivity, $\kappa$, as a
    function of quark volume fraction, $\chi$, for matter at a
    temperature of $10^7$~K and $10^8$~K. The colored lines show the contributions
    of electron-blobs scattering in the mixed quark-hadron phase to $\kappa$,
    computed for parameter sets HV1 (solid) and HV2 (dashed), for
    different surface tensions (see Eq.\ (\ref{eq:surfaceten})):
    $\eta=0.5$ (red), $\eta=1$ (green), $\eta=2$ (blue). The thick black
    lines show thermal conductivities of a standard (no blobs) quark-hadron
    gas computed for HV1 (solid) and HV2 (dashed). The thin black
    lines show the effective bulk thermal conductivities defined in
    Eq.\ (\ref{eq:kappaeff.70}) in the presence of
    rare phase blobs embeded in dominant phase without contribution
    from electron-blob scattering for HV1 (dotted line) and HV2 (dash-dotted line).}
\label{fkappa1}
\end{figure}
lead to large uncertainties in the total neutrino emissivity, since
the neutrino emissivities for this temperature-density regime are
dominated by the modified Urca process and nucleon-nucleon and
quark-quark Bremsstrahlung processes.

\subsection{Thermal conductivity} \label{sec:reskappa}

\begin{figure}[!h]
  \includegraphics[scale=0.58]{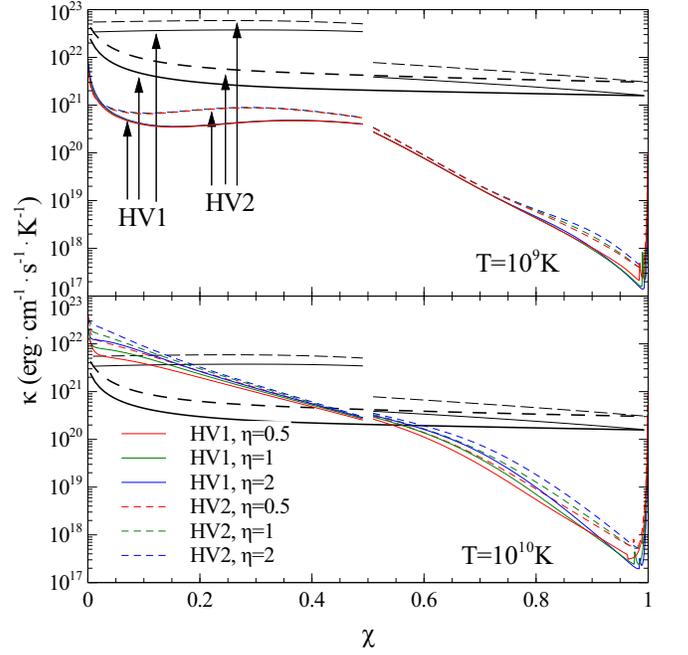}
  \caption{(Color online) Same as Fig.\ \ref{fkappa1}, but for a
    temperature of $T=10^9$~K and $10^{10}$~K.}
\label{fkappa2}
\end{figure}

\begin{figure}[!h]
  \includegraphics[scale=0.58]{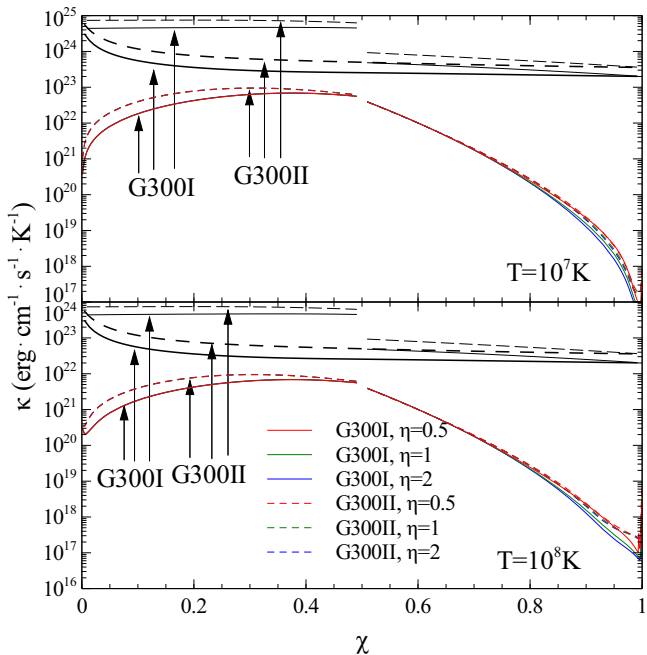}
  \caption{(Color online) Thermal conductivity, $\kappa$, as a
    function of quark volume fraction, $\chi$, for matter at a
    temperature of $10^7$~K and $10^8$~K. The colored lines show the contributions
    of electron-blobs scattering in the mixed quark-hadron phase to $\kappa$,
    computed for parameter sets HV1 (solid) and HV2 (dashed), for
    different surface tensions (see Eq.\ (\ref{eq:surfaceten})):
    $\eta=0.5$ (red), $\eta=1$ (green), $\eta=2$ (blue). The thick black
    lines show thermal conductivities of a standard (no blobs) quark-hadron
    gas computed for G300I (solid) and G300II (dashed). The thin black
    lines show the effective bulk thermal conductivities defined in
    Eq.\ (\ref{eq:kappaeff.70}) in the presence of
    rare phase blobs embeded in dominant phase without contribution
    from electron-blob scattering for G300I (solid) and G300II (dashed).}
\label{fkappa3}
\end{figure}

\begin{figure}[!h]
  \includegraphics[scale=0.58]{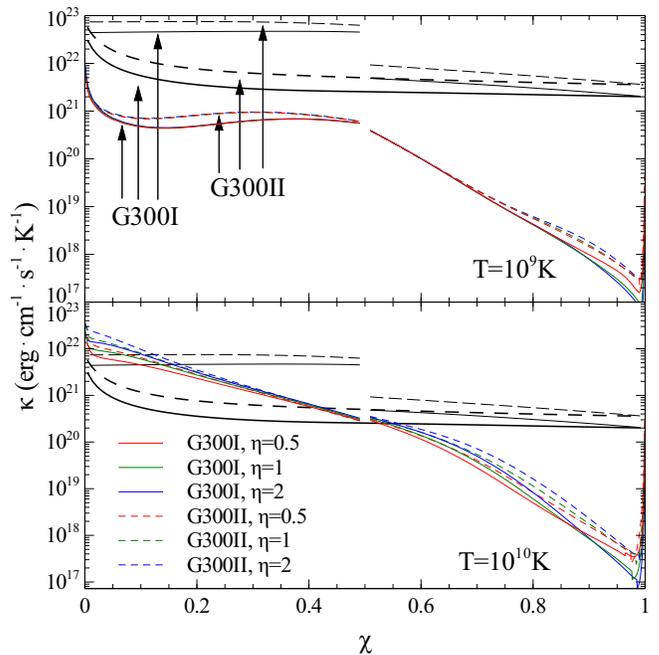}
  \caption{(Color online) Same as Fig.\ \ref{fkappa3}, but for a
    temperature of $T=10^9$~K and $10^{10}$~K.}
\label{fkappa4}
\end{figure}

Figures~\ref{fkappa1} through \ref{fkappa4} show the thermal
conductivities due to presence of rare phase blobs. For comparison,
thermal conductivities in standard quark-hadron gas are shown too
where the thermal conductivity for hadronic phase $\kappa_{\text{H}}$
is adopted from \cite{flowers81:a} and that of quark phase $\kappa_{\text{Q}}$
is adopted from \cite{haensel91:a}. The total thermal conductivity in
standard quark-hadron gas without blobs is given by
\begin{equation}
  \kappa=
  \left(\frac{\chi}{\kappa_{\text{Q}}} + 
  \frac{1-\chi}{\kappa_{\text{H}}} \right)^{-1} \,,
\label{eq:kappaeff}
\end{equation}
Besides scattering between electron and rare phase blobs calculated
in \S \ref{sec:kappa}, the geometric pattern will also alter
the total thermal conductivity.
The two contributions $\kappa_{\text{H}}$ and $\kappa_{\text{Q}}$ from standard quark-hadron gas
can be combined using an expression for the total effective thermal conductivity of spheres
immersed in continuous matter of a different thermal conductivity
\cite{bird02:book}. In our case the bulk thermal conductivity of two phases can be written as
\begin{equation}
  \kappa_{\text{eff}}=
  \kappa_1\left(1-\frac{3\chi}{(2\kappa_1+\kappa_2) /
      (\kappa_1-\kappa_2)+\chi}\right) \,,
\label{eq:kappaeff.70}
\end{equation}
where $\kappa_1$ and $\kappa_2$ are the thermal conductivities of the
dominant and the rare phase, respectively.

The small jumps of thermal conductivities from electron-blob scattering (color lines
in Figs.\ \ref{fkappa1} through \ref{fkappa4}) are due to discontinuities
of rare phase blob mass $m_b$ (See Fig.\ \ref{fig:blobA}). The jumps of
thermal conductivities contributed by embedding of rare phase blobs (thin black lines in
Figs.\ \ref{fkappa1} through \ref{fkappa4}) are due to unequal
thermal conductivities of the two phases $\kappa_1\neq \kappa_2$ at $\chi=0.5$.

As can be seen from Figs.~\ref{fkappa1} through \ref{fkappa4}, the
total thermal conductivity $\kappa = (\kappa_{\text{eff}}^{-1} +
\kappa_{\text{blob}}^{-1})^{-1}$ is dominated by electron-blob
scattering at $T\lesssim 10^9K$. This is particularly the case for
quark volume fractions close to $\chi=0$ or $\chi=1$ where the
electron thermal conductivity from blob scattering can be as much as
three (for $\chi<0.5$) to six orders of magnitude
($\chi>0.5$) smaller than the contribution from the mixed quark-hadron
phase. Physically, this causes a blocking of the thermal flow through
a mixed quark-hadron phase region, which could manifest itself in the
thermal evolution of quark-hybrid stars.

As mentioned in Section~\ref{sec:mixedph}, the blob volume density
$n_b\to 0$ as $\chi\to 0$ or $\chi\to 1$.  Since the Coulomb logarithm
$\Lambda_\kappa$ is finite for these limits but the coefficient
$\propto Z_{\text{eff}}$ is divergent, the thermal conductivity
stemming from rare phase blob scattering will diverge on both ends of the
quark-hadron phase, leading to a vanishing total thermal
conductivities there. Since this
occurs only very near the edges of the quark-hadron boundary
($\chi\lesssim 10^{-2}$) this feature can not be seen in
Fig.~\ref{fkappa1}. The small jumps in $\kappa$ near the edges of the
quark-hadron phase for $T=10^9$K and $10^{10}$K (see
Figs.~\ref{fkappa2} and \ref{fkappa4}) are due to melting.

\section{Summary and Conclusions} \label{sec:conclusion}

Because of the competition between the Coulomb and the surface
energies associated with the positively charged regions of nuclear
matter and negatively charged regions of quark matter, the mixed phase
may develop geometrical structures (e.g., blobs, rods, slabs),
similarly to what is expected of the sub-nuclear liquid-gas phase
transition.  In this paper we explore the consequences of a Coulomb
lattice made of rare phase blobs for the thermal and transport properties
of neutron stars. The total specific heat, $c_V$, thermal
conductivity, $\kappa$, and electron-blob Bremsstrahlung neutrino
emissivities, $\epsilon_{\nu,\text{BR}}$, are calculated and compared
with those of standard neutron star matter.  To carry out this
project, we have adopted, and expanded on, methods of earlier works on
the transport properties of neutron stars
\cite{glendenning89:a,glendenning01:a}.  The sizes of, and spacings
between, rare phase blobs are calculated using the Wigner-Seitz
approximation \cite{glendenning01:a}.  The equations of state used in
this study are computed for a standard non-linear nuclear Lagrangian,
and the associated equations of motion for the baryon and meson fields
are solved in the relativistic mean-field approximation. Quark matter
has been modeled in the framework of the MIT bag model.  Four
different parameter sets (HV1, HV2, G300I, G300II) have been used to
model the composition of neutron star matter containing a mixed phase
of quarks and hadrons (quark-hybrid matter).

The results discussed in Section~\ref{sec:result} show that the
contribution of rare phase blobs in the mixed phase to the specific
heat is negligible compared to the specific heat of a quark-hadron
gas. This is very different for the transport properties.  For low
temperature $T\lesssim 10^8$~K the neutrino emissivity from
electron-blob Bremsstrahlung scattering is at least as important as
the total contribution from other Bremsstrahlung processes (such as
nucleon-nucleon and quark-quark Bremsstrahlung) and modified nucleon
and quark Urca processes (see Figs.\ \ref{fQHV} and \ref{fQG300}).  It
is also worth noting that the scattering of degenerate electrons off
rare phase blobs in the mixed phase region lowers the thermal
conductivity by several orders of magnitude compared to a quark-hadron
phase without geometric patterns (see Figs.\ \ref{fkappa1} through
\ref{fkappa4}).  This may lead to significant changes in the thermal
evolution of the neutron stars containing solid quark-hadron cores,
which will be part of a future study. Another very interesting issue
concerns the impact of more complex geometrical structures (rods and
slabs) on the thermal conductivity and on neutrino transport. The
presence of such structures may reduce the neutrino emissivities
because of changes in the dimension of the reciprocal lattice and the
Debye-Waller factor \cite{kaminker99:a}.

In summary, our study has shown that the presence of rare phase blobs in
dense neutron star matter may have very important consequences for the
total neutrino emissivity and thermal conductivity of such matter. The
implications of this for the thermal evolution of neutron stars need
to be explored in future studies. To accomplish this we intend on
performing two dimensional cooling simulations, in which rotation and a dynamic
composition might be accounted for \cite{Negreiros:2011a,Negreiros:2012a}. In
this connection we refer to the
recent study of Noda {\it et al.}\ \cite{noda11}, who suggested that
the rapid cooling of the neutron star in Cassiopeia A can be explained
by the existence of a mixed quark-hadron phase in the center of this
object.

\section*{Acknowledgments}

The work of X.N. is supported by China Scholarship Council (CSC), and
F.W. is supported by the National Science Foundation (USA) under Grant
No.~PHY-0854699.

\end{document}